\documentclass[final,5p,times,twocolumn]{elsarticle}

\usepackage{amssymb}

\usepackage{subcaption} 
\usepackage{url} 

\journal{NIMA}

\begin{document}

\begin{frontmatter}



\title{A rotor-based multileaf collimator for beam shaping}


\author[inst1]{N. Majernik}

\ead{Majernik@slac.stanford.edu}

\author[inst2]{G. Andonian}
\author[inst2]{A. Parrack}
\author[inst2]{J. B. Rosenzweig}
\author[inst3]{S. Doran}
\author[inst3]{E. Wisniewski}
\author[inst3]{J. Power}

\affiliation[inst1]{organization={SLAC National Accelerator Laboratory},
            city={Menlo Park},
            postcode={94025}, 
            state={CA},
            country={USA}}

\affiliation[inst2]{organization={University of California Los Angeles},
            city={Los Angeles},
            postcode={90095}, 
            state={CA},
            country={USA}}

\affiliation[inst3]{organization={Argonne National Laboratory},
            city={Lemont},
            postcode={60439}, 
            state={IL},
            country={USA}}

\begin{abstract}
We introduce a new style of multileaf collimator which employs rotors with angularly dependent radius to control the masking aperture: a rotor-based multileaf collimator (RMLC). 
Using a padlock-inspired mechanism, a single motor can set dozens of rotors, i.e. leaves, independently. 
This is especially important for an ultra-high vacuum (UHV) compatible MLC, since this reduces the number of actuators and vacuum feedthroughs required by more than an order of magnitude. 
This new RMLC will complement previous work employing a UHV compatible MLC with an emittance exchange beamline to create arbitrarily shaped beams on demand.
A feed-forward control system which abstracts away the complexity of the RMLC operation, and is adaptable to real beamline conditions, is discussed and demonstrated in simulation.
\end{abstract}




\end{frontmatter}


\section{Introduction}
\label{introduction}

A multileaf collimator (MLC) is a device with dozens of independently actuated inserts, or leaves, which intercept or scatter a beam, creating a custom aperture \cite{jordan1994design,boyer1992clinical,ge2014toward}. 
MLCs are common in radiotherapy applications to shape the radiation beam to precisely match the shape of the target from any angle, thereby delivering an effective dose while reducing exposure to non-targeted regions. MLCs can also be used to shape beams for accelerator physics applications. We have previously demonstrated an ultrahigh vacuum (UHV) compatible MLC with forty, 2 mm wide tungsten leaves, each of which was independently magnetically coupled to the exterior of the vacuum vessel and actuated with a 3D printed drivetrain module \cite{mlcV1}. It was used in conjunction with an emittance exchange (EEX) beamline to rapidly create near-arbitrary beam current profiles. The MLC replaced the previously employed laser-cut tungsten masks and enabled real-time control over the beam profile with the same fidelity as the laser-cut masks.

Here, we introduce a new style of multileaf collimator based on rotors with angularly dependent radius. This new design reduces the number of vacuum feedthroughs and actuators by an order of magnitude, dramatically simplifying the mechanical design. Further, this new design scales more favorably to larger aperture, permitting any anticipated beam through unobstructed, allowing the new MLC to remain permanently installed on the beamline, providing a new capability to users. We will also discuss a new feed-forward control scheme to enhance the performance of the MLC.

\begin{figure*}[t]
   \centering
   \includegraphics[width=\textwidth]{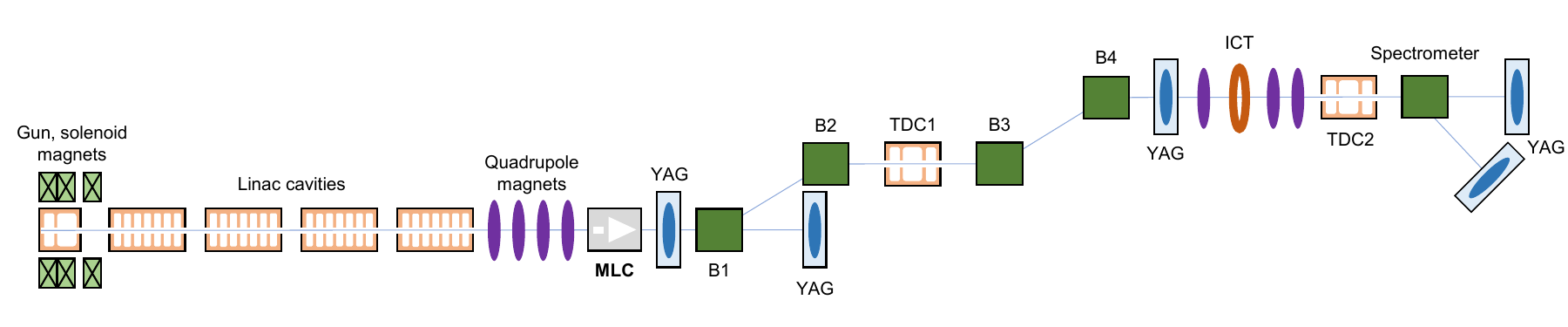}
   \caption{AWA drive linac and EEX beamline (not to scale), adapted from~\cite{ha2017precision}.}
   \label{fig:beamlineSchematic}
\end{figure*}

In modern accelerator physics, one of the goals is the full control of particle beam distributions in multi-dimensional phase spaces \cite{Ha:2022rev}.
While methods exist for transverse phase space shaping, \textit{e.g.} employing magnetic elements or rigid collimators, there are fewer reliable options for longitudinal phase space tailoring.
Optimally designed longitudinal profiles of beam current are important in many applications. For example, asymmetric (ramped) beam profiles can dramatically enhance efficiency in wakefield-driven acceleration concepts \cite{bane1985collinear,lemery2015tailored}, mitigate effects stemming from coherent synchrotron radiation \cite{Mitchell:2013}, and enhance the final energy output in free-electron lasers \cite{Ding:2016}.
Specifically, transverse distribution masking combined with EEX \cite{Sun:PhysRevLett.105.234801,PhysRevAccelBeams.19.121301} (See Fig. \ref{fig:beamlineSchematic}) is a versatile, high precision option for shaping the longitudinal profiles of high charge bunches. 
In EEX, one of the transverse phase-space planes of the beam is swapped with the longitudinal phase plane, for example, by placing a transverse-deflecting cavity between two dogleg transport sections~\cite{Emma:PhysRevSTAB.9.100702}; similar shaping of high charge current profiles is difficult to achieve using other techniques \cite{Ha:2022rev}. 
Replacing a fixed mask with an in-vacuum MLC, permitting real time tuning during experimental operations, opens up new possibilities. These notably include online optimization using machine-learning methods, which give the opportunity to refine not only the delivered beam profile, but also the performance of the final (\textit{e.g.} wakefield accelerator) system. 

The RMLC is used at the the AWA EEX beamline, a linear accelerator which accelerates multi-nanocoulomb electron bunches up to 43 MeV energy.
A series of quadrupole magnets are used to control the beam transverse-phase-space parameters at the location of the transverse mask. 
After the beam is masked, it traverses the EEX beamline, which consists of four dipoles and a transverse-deflecting cavity (TDC) to swap the horizontal and longitudinal phase spaces. The longitudinal phase space downstream of the EEX beamline is characterized in a diagnostic line, as illustrated in  Fig.~\ref{fig:beamlineSchematic}.

\section{Rotor-based multileaf collimator}

\begin{figure}[h]
   \centering
   \includegraphics*[width=1.0\columnwidth]{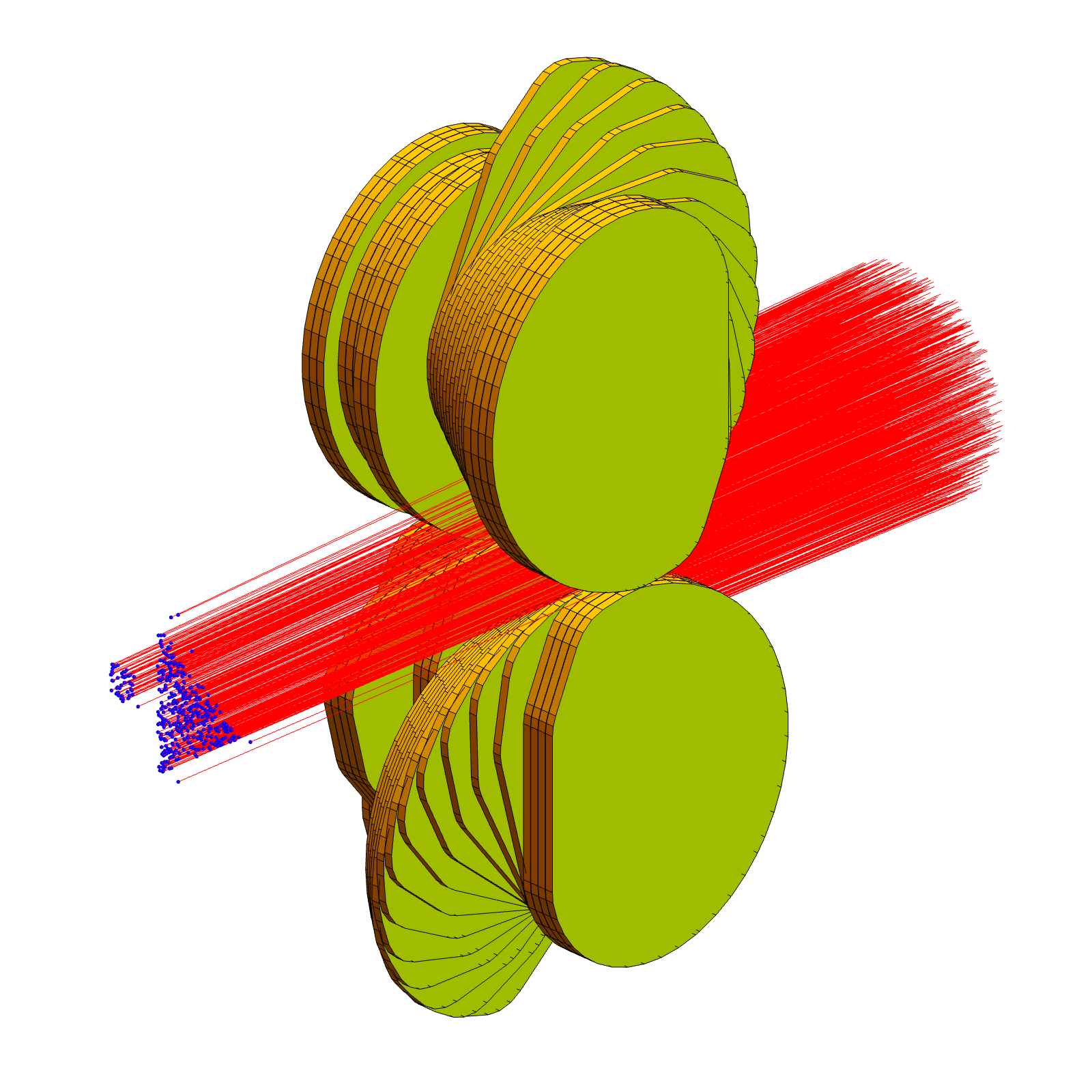}
   \caption{Simplified schematic of a RMLC masking a beam (traveling from right to left).}
   \label{fig:schematicMasking}
\end{figure}

A simplified schematic of an RMLC masking a beam is shown in Figure \ref{fig:schematicMasking}, illustrating how, by changing the angle of each rotor, the depth to which it intercepts the beam is controlled. This design is more favorable, from a practical perspective, for scaling to a larger aperture. The old MLC maximum opening was 20 by 40 mm \cite{mlcV1}, this new one is more than 50 by 50 mm, which is large enough to let any anticipated AWA beam through unobstructed. This means that the new MLC can be left permanently installed on the beamline, providing a new capability for AWA users. The overall height of the previous style of the MLC scales as the product of the number of leaves and the length of the leaf throw: to achieve the same aperture with the previous design, the MLC would be infeasibly large.

\begin{figure}[htbp]
   \centering
   \includegraphics*[width=1.0\columnwidth]{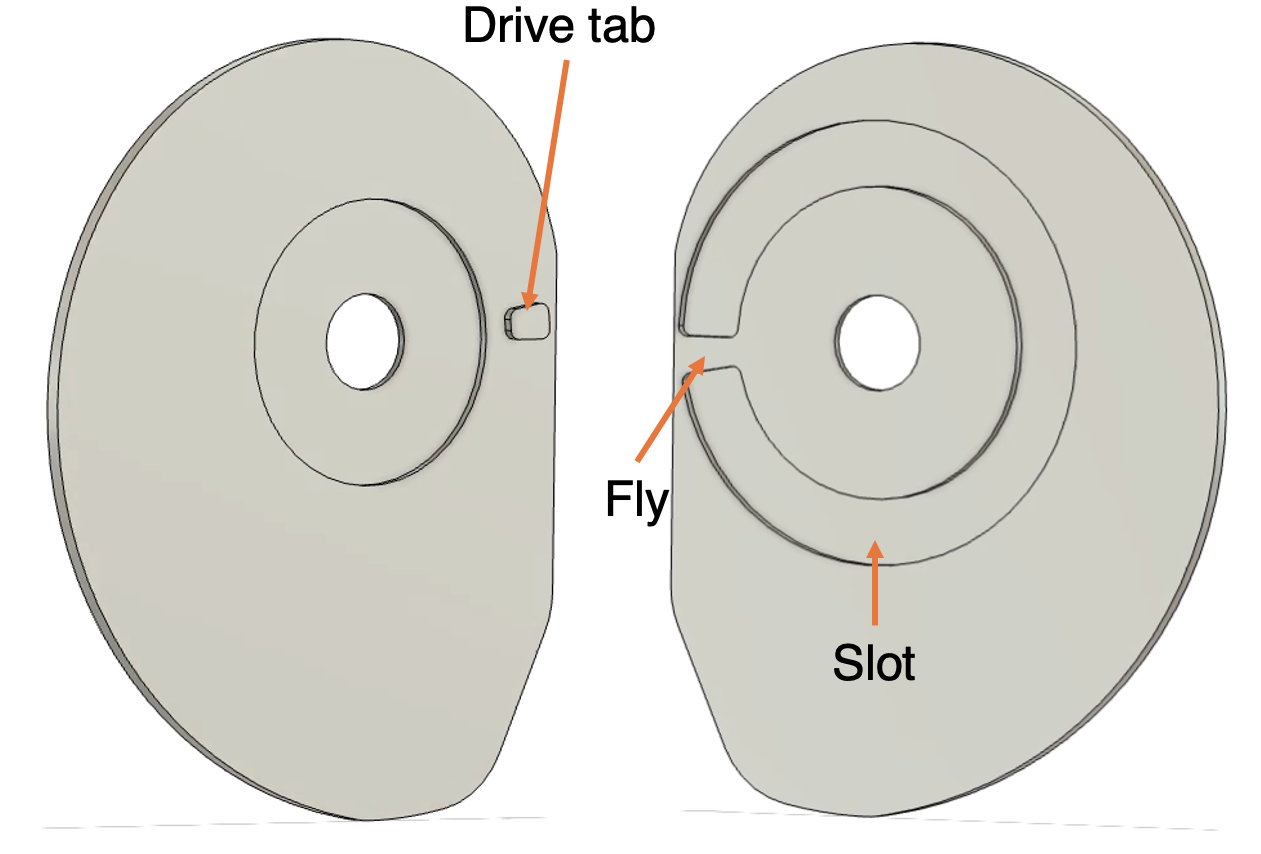}
   \caption{Render showing the geometry of both faces of the rotor. Key elements are annotated.}
   \label{fig:rotorRender}
\end{figure}

\begin{figure*}[t]
   \centering
   \includegraphics[width=0.75\textwidth]{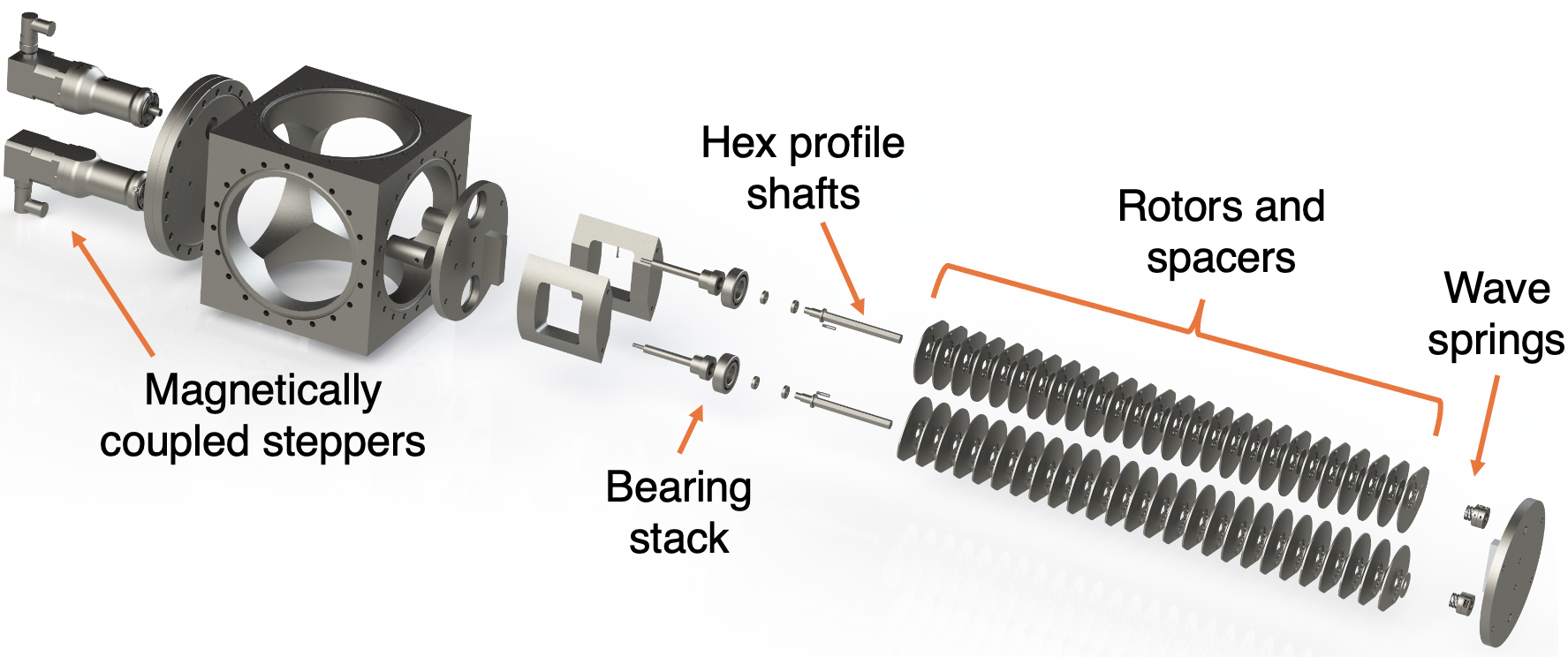}
   \caption{Exploded render of the RMLC engineering design.}
   \label{fig:explodedRender}
\end{figure*}

The actual rotor design, fabricated from 304 stainless steel, is shown in Figure \ref{fig:rotorRender}, which includes tabs and slots. When these rotors are stacked together, the \textit{drive tab} sits in the slot, and is able to turn freely without touching the adjacent rotor, until it contacts a tab in the slot (the \textit{fly}). Once the drive tab and fly are in contact, as long as the rotation continues in the same direction, the two rotors move together. This process repeats, and rotors can be ``picked up'' one-by-one, until the whole stack is rotating as a unit. Then, rotation continues until the last rotor is at the desired angle. At this point, the direction of rotation is reversed and the driven rotor loses contact with the second rotor. Eventually the drive tab makes contact with the opposite side of the second rotor's fly, and the second rotor begins to rotate. This process repeats, until the second to last rotor is engaged and moved to its desired angle. This alternation repeats, setting the rotors one by one from the far side to the driven side. This is analogous to how a combination padlock or bank safe works \cite{blaze2004safecracking} (See [supplementary video]). Crucially, this means only a single actuator and vacuum feedthrough per stack are required, dramatically simplifying the mechanical and electronic complexity of the RMLC.

\begin{figure}[tb]
   \centering
   \includegraphics*[width=1.0\columnwidth]{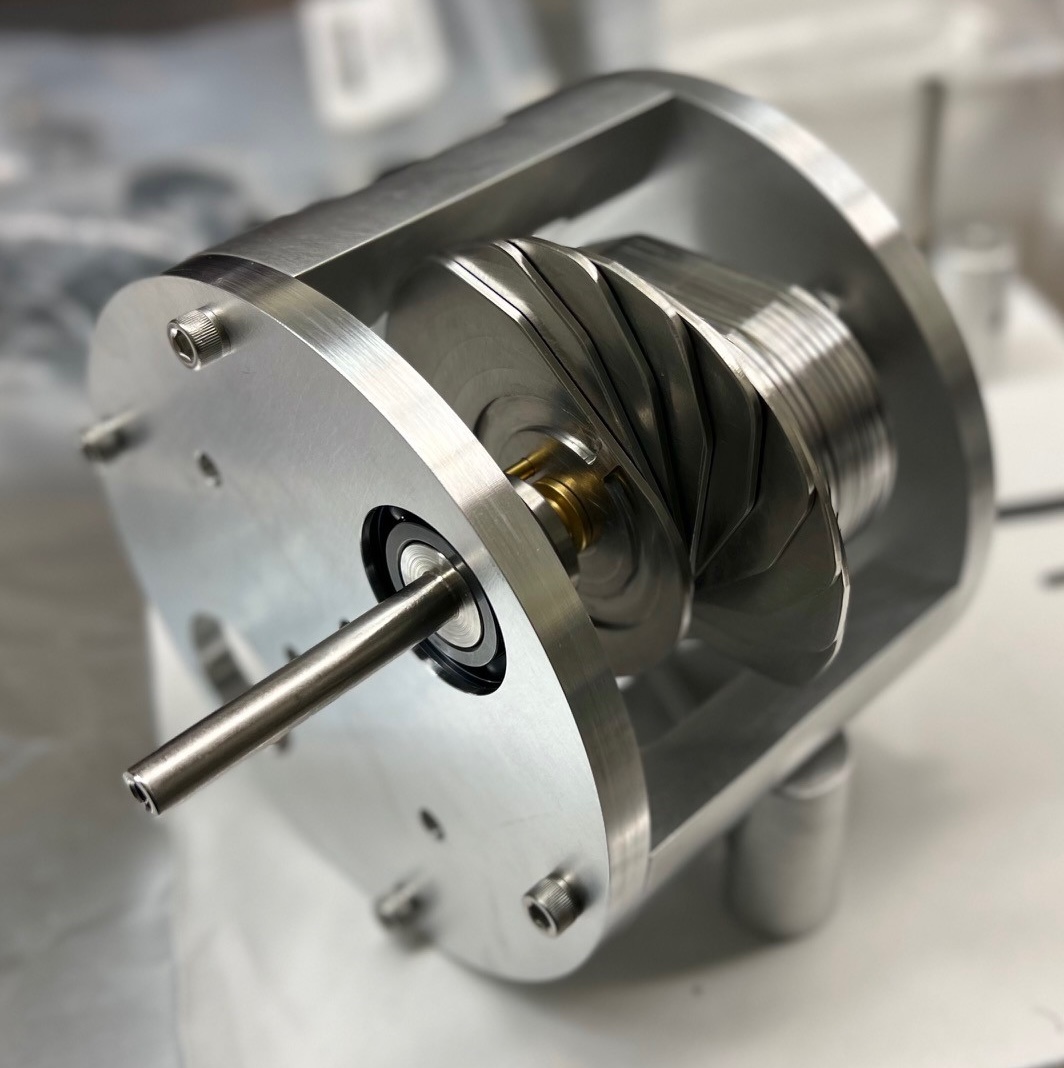}
   \caption{Partially assembled RMLC.}
   \label{fig:posedPhoto}
\end{figure}

To mask the beam, two of these rotor stacks will be employed, one above and one below the beam, each driven by a commercially available, magnetically coupled rotary feedthrough. The final design (Figure \ref{fig:explodedRender}) fits inside an 8” cube and is designed to be completely assembled outside the cube, mounted on a conflat flange, and installed as a unit. Between each pair of rotors is a static spacer. The center shaft has a hexagonal profile which matches a cutout in the spacer so they are unable to rotate while the rotors have a large, circular hole which rides on fillets on the hex profile corners. The whole stack of rotors and spacers is axially compressed with a wave spring with adjustable force. This ensures that the rotors are held in place so they do not change angle, except when the drive tab and fly are in contact. To maintain consistent friction between each set of contact areas, the spacers and the hex shaft have a titanium nitride coating. The partially assembled RMLC is shown in Figure \ref{fig:posedPhoto}.

In contrast with the laser cut masks and previously employed MLC, which functionally masked in a 2D plane, the RMLC will interact with and mask a non-collimated beam in a three dimensional fashion. We've established a simulation pipeline using OPAL \cite{adelmann2009object} which treats the RMLC as a series of 2D mask elements to account for this behavior. To minimize leakage between rotors, the RMLC can be installed with a 2 degree rotation about the vertical axis; this removes line-of-sight between the rotors but negligibly impacts the RMLC resolution.

\section{Feed-forward control}
To control the new RMLC, we have also developed a new, feed-forward control scheme and demonstrated its operation in simulation. Previously, masks were predetermined by simulation: a desired current profile was specified and OPAL simulations of the beamline would define the mask. However, an accelerator's behavior can never be fully captured in simulation so a mask that is informed by the real-world accelerator conditions will give superior results. Additionally, this approach will enable the mask to be updated to respond to accelerator drifts, something not achievable with the simulation-defined masks. Further, this new system abstracts away much of the complexity of the RMLC from the user including the 3D nature of the masking: it is only necessary to specify a desired current profile (or other objective) and the control system can appropriately set the RMLC.

The free parameters are the $n$ leaf positions and the output is a current profile after EEX, described by the current in $k$ bins along the beam's longitudinal coordinate, $\zeta$. The process takes place in two steps: an initial guess, followed by refinement. For the first stage of this process where we make an initial guess, we neglect collective effects and so treat the current contributions from each leaf as perfectly additive.

\begin{figure}[t]
   \centering
   \includegraphics*[width=1.0\columnwidth]{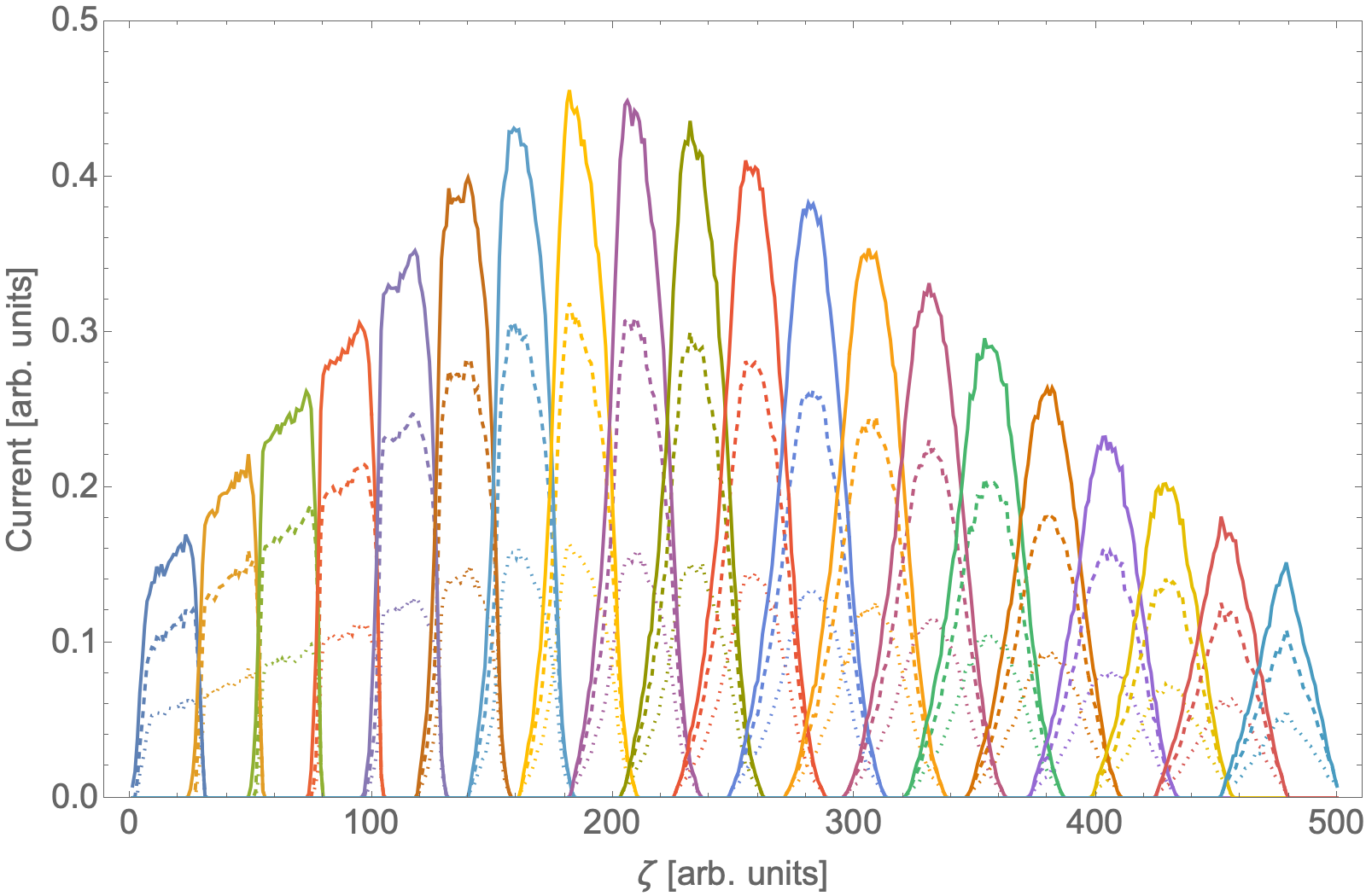}
   \caption{Response functions for some of the RMLC leaves. Each color corresponds to a single leaf. The dotted line is 33\% opening, the dashed line is 66\% opening, and the solid line is completely open.}
   \label{fig:responseFunctions}
\end{figure}

The first step is to experimentally determine the response functions of each of the leaves. The position, $x_i$, of the $i^{th}$ leaf will correspond to a vector in the output space, described by the response function, $\vec{f_i}(x_i)$. Naively, these response functions may be determined by closing all other leaves, setting one leaf to several positions, and interpolating. More efficient methods allow the response functions to be obtained experimentally, more rapidly. Since, for this initial guess, collective effects are ignored, we assume that the resultant current vector is simply the sum of all the leaf contributions, $\vec{I}(\vec{x}) = \sum_{i=1}^{n} \vec{f_i}(x_i)$. The goal is to find the leaf positions that minimize the distance between this sum and the user's desired current profile, $\vec{I_t}$. To demonstrate the method, we've applied it to an OPAL simulation of the AWA beamline, recovering the response functions for each of the leaves (Figure \ref{fig:responseFunctions}). The smearing arising from the EEX process causes adjacent leaves to have overlapping current contributions.

Under the assumptions described, increasing the opening of a leaf will monotonically increase the current in all bins. Unfortunately though, the effect of increasing the opening is not a straight line in the $k$-dimensional output space; this can be observed in Figure \ref{fig:responseFunctions} by how the curves change shape (beyond simply scaling) at different opening fractions. This rules out the use of some convenient optimization strategies like maximum likelihood estimation. Instead, traditional numerical optimization has been employed, which doesn't require this linearity, albeit at the cost of greater computational expense. 

Once we have an initial guess for the mask, we move on to refinement. The MLC is set to this initial guess, then each of the leaves is perturbed from its nominal position and a new response function is inferred. For this step, we now assume a linear response since we expect that our changes from the nominal state are small. This has the added benefit of holding the impact of collective effects roughly constant. With these new response functions in hand, we re-run the optimizer to find the new leaf positions.

\begin{figure}[t]
   \centering
   \begin{subfigure}[b]{1.0\columnwidth}
       \centering
       \includegraphics[width=1.0\columnwidth]{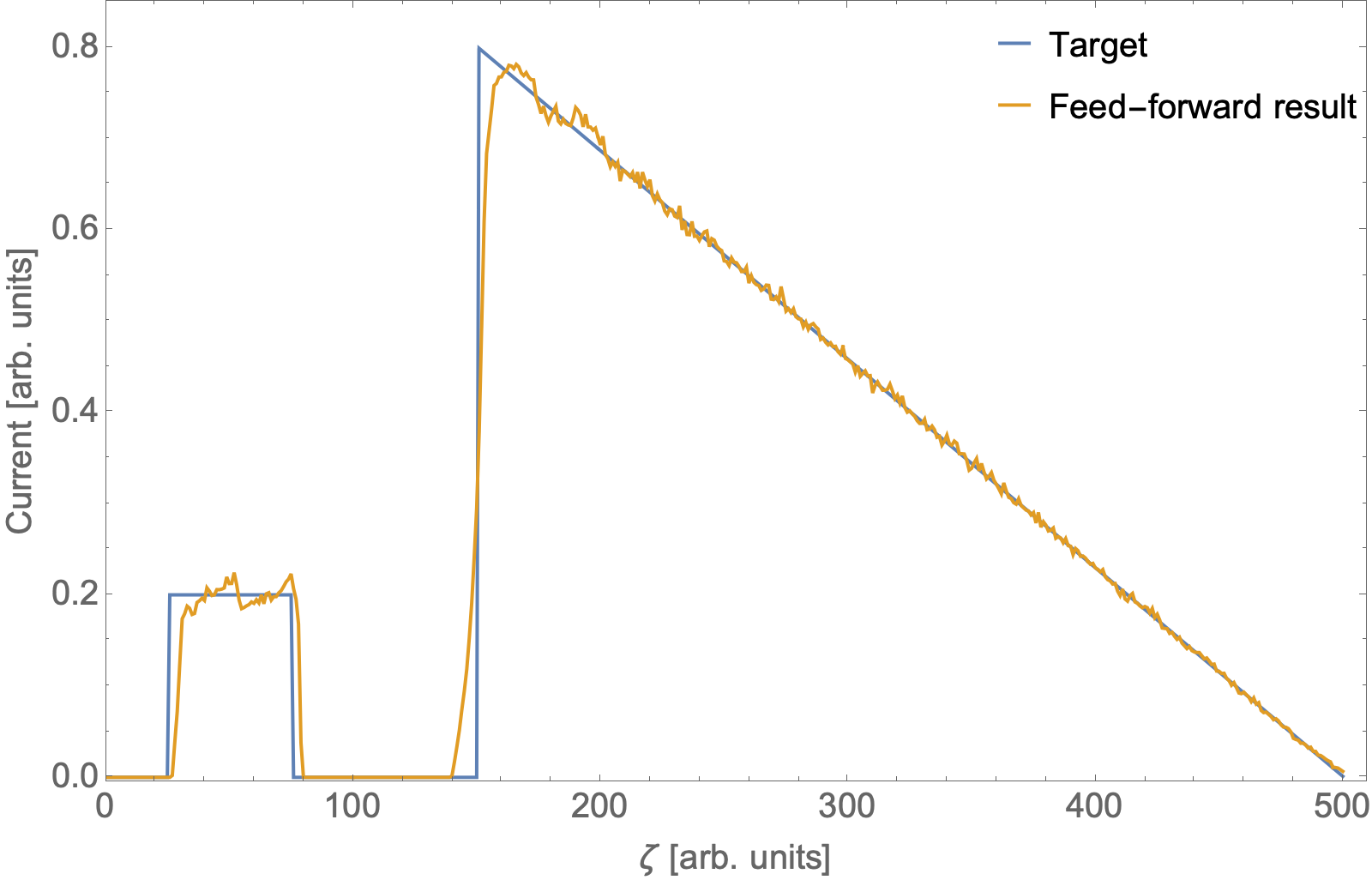}
       \caption{User specified current profile in blue and the best resultant profile in orange.}
       \label{fig:resultCurrent}
   \end{subfigure}
   
   \vspace{0.5cm}
   
   \begin{subfigure}[b]{1.0\columnwidth}
       \centering
       \includegraphics[width=1.0\columnwidth]{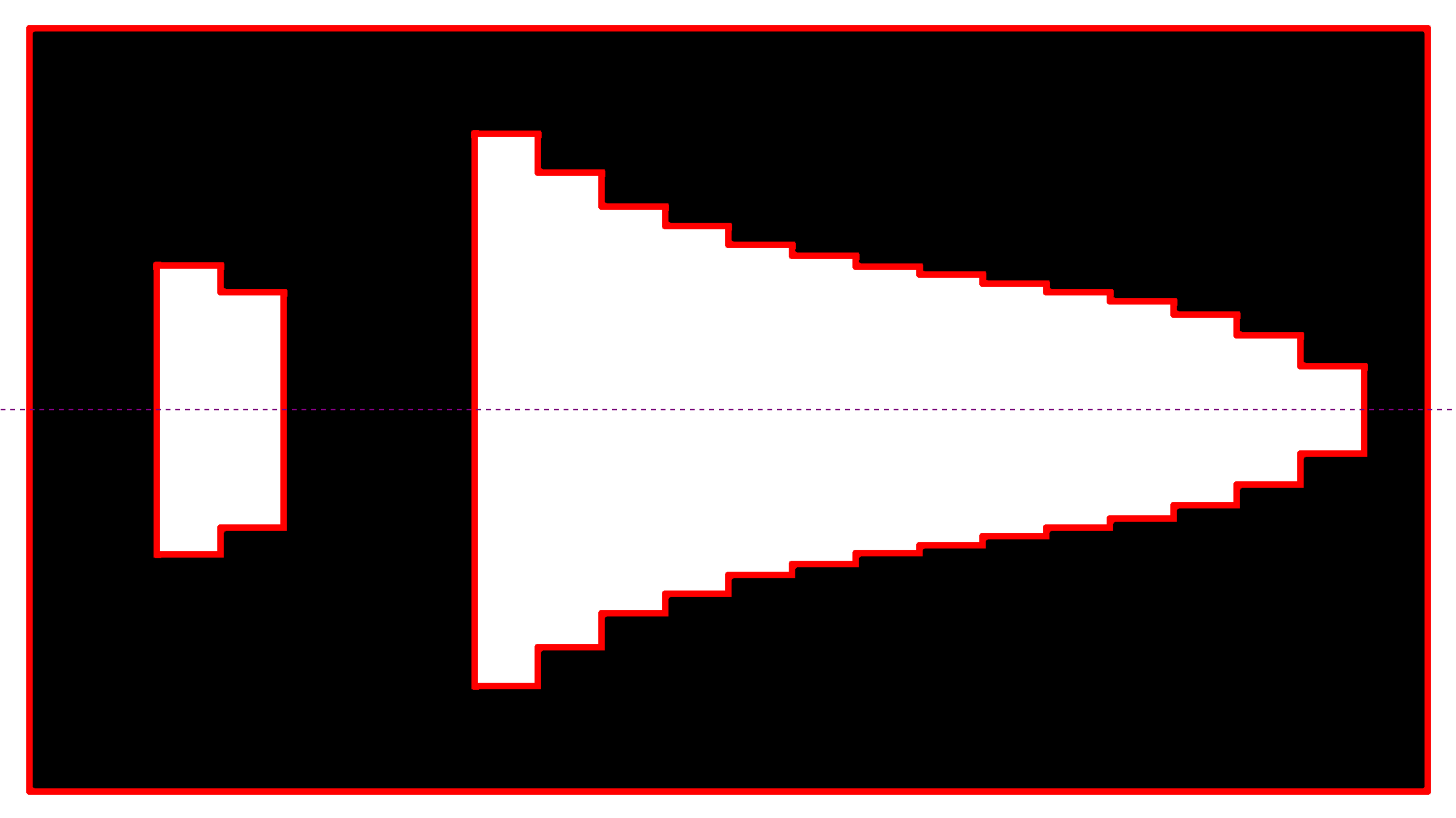}
       \caption{The feed-forward optimized mask.}
       \label{fig:resultMask}
   \end{subfigure}
   
   \caption{Simulated feed-forward controller results.}
   \label{fig:resultCombined}
\end{figure}

Figure \ref{fig:resultCombined} shows the result of this feed-forward optimization process in an OPAL simulation. A user specified current profile, a linearly ramped drive beam and trailing witness, is defined and the optimizer attempts to set the leaves to achieve it. The mask that produces this current profile is shown in Figure \ref{fig:resultMask}. For this particular optimization, a symmetry condition is imposed, requiring upper and lower leaves to have equal openings. Naturally, the objective function can be modified to impose other constraints. This same optimization process can also apply to objectives which are not current profiles but still adhere to the assumptions about being able to sum contributions from each leaf.

\section{Summary and conclusions}
We have developed a rotor-based multileaf collimator for masking beams. This particular instance of an RMLC operates on an ultra-high vacuum linac for advanced accelerator research. By employing a padlock-inspired mechanism, RMLC can be set using only two actuators and two motion feedthroughs, while maintaining UHV compatibility. This new design scales to larger aperture and so can remain permanently installed on a beamline. A new feed-forward control scheme abstracts away the complexity of the device for users, allowing them to simply specify a desired current profile, and having the mask set according to the current accelerator state. Since the accelerator conditions can be directly measured with the RMLC, it will be able to produce these current profiles with greater fidelity than simulation-defined masks.

The concept may find use in other accelerator beamlines that rely on transverse masking and require strict UHV levels, for example at the BNL ATF \cite{Muggli:2008} or at SLAC FACET-II \cite{yakimenko2019facet}. 
Since the MLC has a large number of variables for tuning, it is an ideal instrument to pair with machine-learning-based control systems \cite{scheinker2021extremum, duris2020bayesian}. This approach would enable many applications, including high transformer ratio wakefield acceleration and improved X-ray FEL performance.

\section{Patents}
N. Majernik, G. Andonian, J. Rosenzweig, “System and methods for multileaf collimator,” U.S. Provisional Patent US 63/422,691, Nov. 2022.

\section{Acknowledgments}
This work was supported by the National Science Foundation under Grant No. PHY-1549132, DOE Grant No. DE-SC0017648, and DOE Contract No. DE-AC02-06CH11357.

 \bibliographystyle{elsarticle-num} 
 \bibliography{References}





\end{document}